\documentclass[12pt,preprint]{aastex}

\usepackage{verbatim,hyperref}

\shorttitle{}
\shortauthors{}

\begin{document}

\title{Physics GRE Scores of Prize Postdoctoral Fellows in Astronomy}
\author{Emily M.~Levesque$^{1}$, Rachel Bezanson$^{2}$, \& Grant R.~Tremblay$^{3}$}
\affil{$^1$Department of Astronomy, Box 351580, University of Washington, Seattle, WA 98195, USA; \textup{\texttt{emsque@uw.edu}}} 
\affil{$^2$Steward Observatory, University of Arizona, Tucson, AZ 85721, USA; \textup{\texttt{rbezanson@email.arizona.edu}}} 
\affil{$^3$Yale Center for Astronomy \& Astrophysics, Yale University, New Haven, CT 06511, USA; \textup{\texttt{grant.tremblay@yale.edu}}}

\begin{abstract}
The Physics GRE is currently a required element of the graduate admissions process in nearly all U.S. astronomy programs; however, its predictive power and utility as a means of selecting ``successful" applicants has never been examined. We circulated a short questionnaire to 271 people who have held U.S. prize postdoctoral fellowships in astrophysics between 2010-2015, asking them to report their Physics GRE scores (this should not in any way be interpreted as a belief that a prize fellowship is the best or only metric of ``success" in astronomy). The response rate was 64\%, and the responding sample is unbiased with respect to the overall gender distribution of prize fellows. The responses reveal that the Physics GRE scores of prize fellows do not adhere to any minimum percentile score and show no statistically significant correlation with the number of first author papers published. As an example, a Physics GRE percentile cutoff of 60\% would have eliminated 44\% of 2010-2015 U.S. prize postdoctoral fellows, including 60\% of the female fellows. From these data, we find no evidence that the Physics GRE can be used as an effective predictor of ``success" either in or beyond graduate school.
\end{abstract}

\section{Introduction}

The Graduate Record Exam (GRE) subject test in Physics currently plays a significant role in graduate admissions for U.S. astronomy departments. Out of the 20 top-ranked astronomy graduate programs in the U.S. (National Research Council rankings, 2010; excluding physics departments), 100\% of them required the Physics GRE (PGRE) as part of a complete application to their graduate program in 2015. While the role that the PGRE plays in assessment varies among individual departments and admissions committees, applications are often subject to formal or informal ``cutoffs" based on the PGRE percentile score (with a perfect score earning a percentile of 99\%). This practice is based on the assumption that the PGRE is a useful gauge of whether applicants will be ``successful", both as graduate students and as professionals. Lower-scoring applicants are either eliminated from the pool entirely or required to compensate for a low PGRE score with exceptional credentials in other areas of their application. Prospective applicants with low PGRE scores will sometimes choose to not even apply to particular programs - particularly those which quote an ``average" or ``typical" PGRE score on their websites - out of a belief that their score will automatically render them uncompetitive in the applicant pool.

Standardized tests similar to the PGRE have a demonstrated record of performance being skewed by gender and race. The Educational Testing Service, which administers the PGRE as well as the General GRE, has demonstrated that both gender and race impact scores on the General GRE, with lower scores among women and underrepresented minorities (see, for example, Steele \& Aronson 1995, Miller \& Stassun 2014). This same demographic skew has been demonstrated for a broad array of standardized tests, including the PGRE itself\footnote{See \url{http://www.ets.org/Media/Tests/GRE/pdf/gre_0809_factors_2006-07.pdf}} (e.g. Miller 2013). 

The predictive power of such exams for long-term outcomes is also unclear. Quantifying ``success" either in or beyond graduate school is a complex undertaking, and reducing this to a single metric is often considered ineffectual. For example, while Glanz (1996) shows that PGRE scores weakly correlate with average coursework grades for {\it physics} graduate students at Harvard, the same work also argues against heavily weighting the Physics GRE in graduate admissions procedures for physics programs. Glanz (1996) also demonstrates no correlation between coursework performance and other metrics of ``success" in graduate school, such as research productivity.

The role of the PGRE in astronomy graduate admissions - or, more specifically, whether PGRE score correlates with ``success" in the field of astronomy - has not been explicitly studied. Demographic trends in the PGRE are hard to examine within small sample groups, and it is difficult to converge upon a satisfactory and broadly inclusive definition of ``success". However, we believe that even an informal examination of this specific scenario - PGRE scores of ``successful" astronomers - can be useful in considering how the PGRE should be used in astronomy graduate admissions.

Here we present data from a questionnaire sent to 271 individuals who have held prize postdoctoral fellowships in astronomy between 2010-2015. The questionnaire asked recipients to self-report their scores on the PGRE.

\section{Experimental Design} 
\subsection{Questionnaire}
We designed a short (13-question) questionnaire focused specifically on PGRE performance. The questionnaire (Figure 1) asked respondents to report the year that they took the PGRE as well as their raw score, specific percentile scores, and/or percentile range, with percentiles split into ten equal ranges spanning from 0-99\%. 

The questionnaire also asked respondents to report their number of {\it first author} papers, both when they completed their undergraduate degrees and when they completed graduate school. This was primarily done  as a means of roughly quantifying research productivity in graduate school. It also offered a means of examining how publications can potentially impact graduate admissions (for example, whether respondents with low PGRE scores generally had strong undergraduate publication records that would have strengthened their graduate school applications).

We did not collect data on respondents' general GRE scores or grade point averages (GPAs). This was primarily done to keep the questionnaire as short as possible and maximize our potential response rate; however, it is also worth noting that the General GRE has been studied elsewhere (see, for example, Miller \& Stassun 2014), and effectively analyzing GPAs would require collecting substantial additional data on respondents' undergraduate institutions, grading practices, majors, and coursework.

The questionnaire was designed to be sent to individuals who held a national prize postdoctoral fellowship (Hubble, Einstein, NSF, Sagan, and/or Jansky) between 2010 and 2015. As a result of this small potential sample we collected only minimal demographic data (English fluency and the option to identify gender) in order to avoid compromising the anonymity of the respondents. While respondents were asked to identify the specific fellowship or fellowships that they held along with the start year, this was done to assess the completeness of our response rate and the data was not used in any comparative analyses. We also asked respondents to characterize the institution that granted their bachelors' degrees, but that data is not used in the analyses presented here.

All respondents were required to answer a question asking whether they consented to the use of their responses in a short astro-ph publication. For the remainder of this work we include only the data from those who answered ``yes" to this question.

\begin{figure}
\centering
\includegraphics[width=0.45\textwidth]{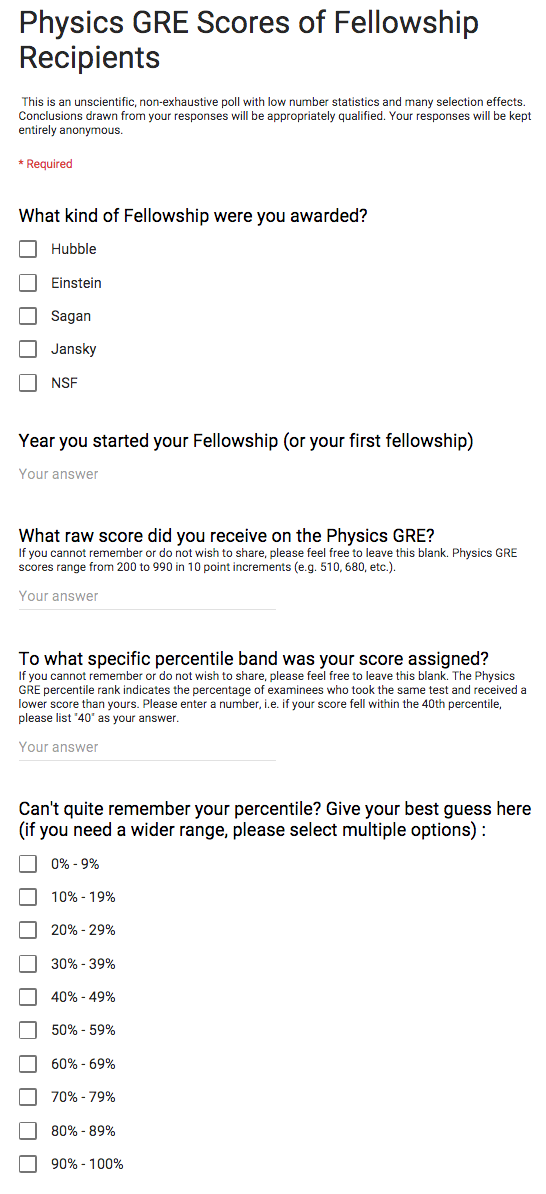}
\includegraphics[width=0.45\textwidth]{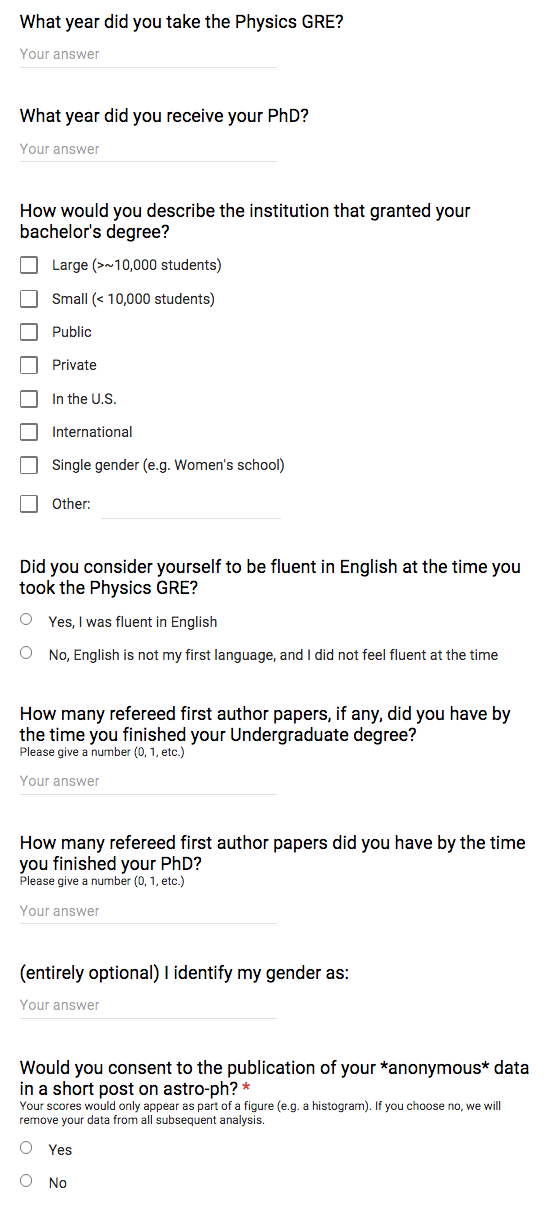}
\caption{Original Questionnaire}
\label{fig:questionnaire}
\end{figure}

\subsection{Responses}
The questionnaire was emailed to 271 potential respondents beginning on 14 November 2015. All of the recipients had held a national prize postdoctoral fellowship (Hubble, Einstein, NSF, Sagan, and/or Jansky) between 2010 and 2015. Responses were collected anonymously; no email addresses or other personal identifications were logged.

Of the 271 questionnaire recipients, 173 responded (response rate of 64\%). 23 responses did not report a PGRE score\footnote{Many prize postdoctoral fellows received their PhDs from non-U.S. institutions that do not require the PGRE.}, leaving us with a total of 149 unique respondents with reported PGRE scores. Of the 149 respondents reporting PGRE scores, 75 identified as male, 53 identified as female, and 21 identified as other or chose not to identify their gender. 

Respondents were given the option of selecting multiple percentile ranges if they could not recall their exact scores or score ranges. 17 people (11\%) gave multiple percentile ranges for their scores. In 3 cases (2\%), these respondents also reported their exact numerical scores and the year that the took the PGRE; this allowed us to assign them to a single percentile bin based on their reported ranges and the score conversions and percentiles published by ETS\footnote{See, for example, the {\it GRE Guide to the Use of Scores} for 2010-2011 available at \url{https://www.ets.org/s/gre/pdf/2010-11_gre_guide.pdf}.}. In the remaining 14 cases (9\%), the median of the reported percentile range was adopted for our analyses.

\subsection{Caveats} 
There are several important caveats worth noting in the design of this questionnaire and study. The small sample size carries with it the inevitable impact of small number statistics. The nature of the questionnaire is also susceptible to the standard errors of self-reporting.

Finally, our distribution of the survey only to prize fellows should not in any way be interpreted as a belief that someone applying to graduate school in astronomy must eventually earn a prize fellowship in order to be considered ``successful". Holding a prize fellowship is a narrow, incomplete, and non-inclusive definition of ``success". Similarly, the number of first author papers published in graduate school is by no means the best or only method of assessing an individual's research productivity; this is particularly true for people working in large collaborations or in fields such as instrumentation or computation.

At the same time, restricting our work to such narrow criteria is useful in a simplistic analysis of whether the PGRE correlates with ``success". Holding a prize postdoctoral fellowship is one (though certainly not the only) marker of ``success" in an astronomy career, and first author publications are one (though not the only) means of assessing ``success" in research. As a result, if the PGRE is indeed an effective and unbiased means of assessing the potential for ``success" in graduate school, then one could reasonably expect that the PGRE scores of prize postdoctoral fellows with strong graduate school publication records should be extremely high.

\section{Results and Analyses}
\begin{figure}[h]
\centering
\includegraphics[width=0.45\textwidth]{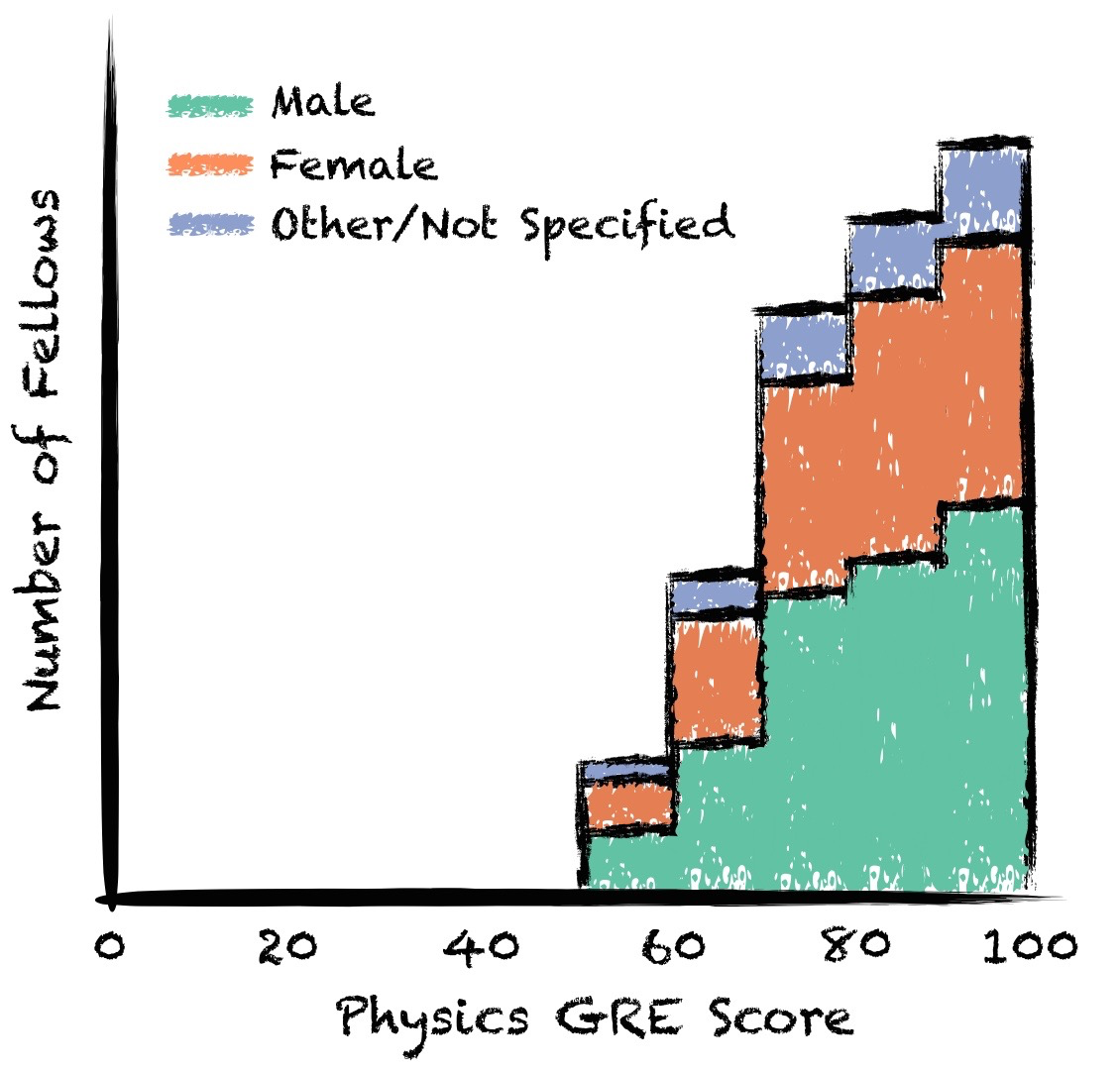}
\includegraphics[width=0.45\textwidth]{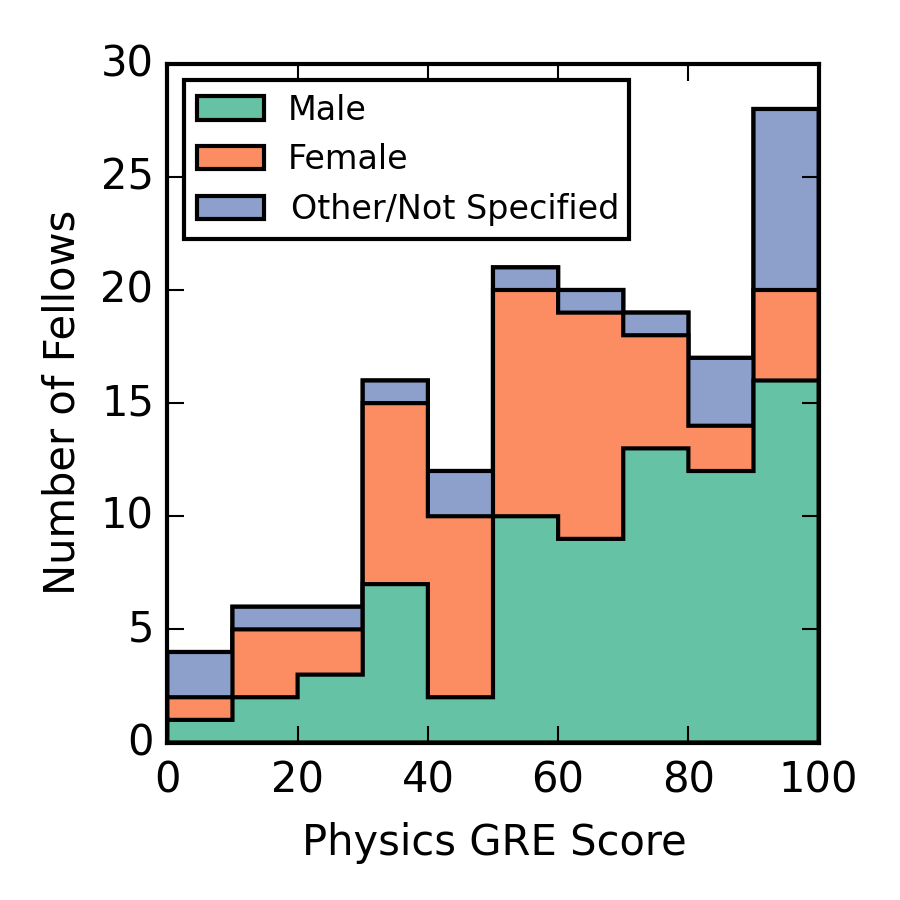}
\caption{{\bf Left}: ``expected" histogram of PGRE percentile scores for respondents if the PGRE was strongly correlated with career performance and a 50th-percentile hard cutoff in graduate admissions was an effective means of selecting future prize fellows. Scores are expected to be distributed as an approximate half-Gaussian, with most fellows falling in the highest bins and no prize fellows with PGRE scores below the graduate admissions cutoff score. This estimate also assumes no gender disparity in PGRE performance; while the gender demographics match those of our survey respondents (50\% male, 35.5\% female, 14\% other/NS), the score distributions are identical. {\bf Right}: Actual histogram of PGRE percentile scores for respondents. The score distribution spans the full range of percentiles, with $\sim$29\% of fellows scoring below the 50th percentile. The distribution of scores is also notably different for male and female fellows; $\sim$42\% of female fellows fall below the 50th percentile.}
\label{fig:gre_hist}
\end{figure}

Figure 2 illustrates the difference between the ``expected" PGRE scores of prize fellows and those found in our data. On the left, we show the expected appearance of a score histogram for prize fellows if the PGRE was strongly correlated with ``success" and a 50th-percentile hard cutoff in graduate admissions was an effective means of selecting potentially successful applicants. In this scenario we would expect scores to be distributed as an approximate half-Gaussian, with most fellows' scores falling into the highest bins and {\it no} prize fellows with PGRE scores below the graduate admissions cutoff score. This estimate also assumes that there is no gender disparity in PGRE performance; while the gender demographics match those of our survey respondents, the score distributions within each gender category are identical.

On the right we show our actual histogram of PGRE percentile scores reported by prize fellows. The score distribution spans the full range of percentiles, with 44 fellows ($\sim$30\%) falling below the 50th percentile. The distribution of scores is also notably different for male and female fellows; $\sim$42\% of female fellows fall below the 50th percentile. The results show a clear disparity in gender performance on the Physics GRE among prize fellows. Multiple fellowships are represented in each percentile bin.

Figure 3 illustrates the number (left) and fraction (right) of fellows - both in total and divided by gender -  that would be eliminated from a theoretical graduate school applicant pool by applying hard thresholds in the PGRE percentiles, with thresholds ranging from 10-60\%. Table 1 offers a numerical breakdown of these values, with thresholds ranging from 10-90\%.

\begin{figure}[h]
\centering
\includegraphics{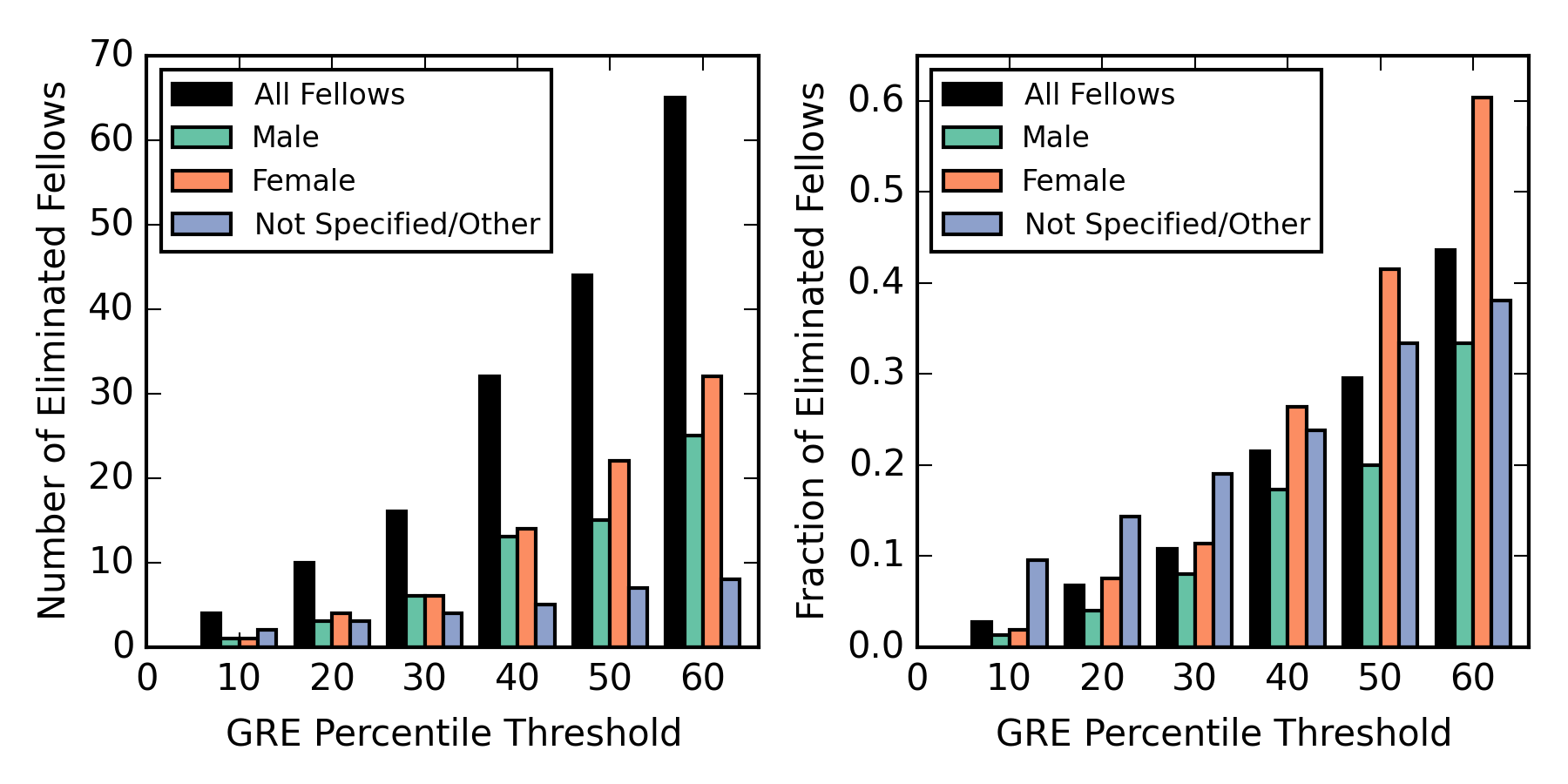}
\caption{Number of Fellows (left panel) and fraction of fellows (right panel) eliminated by hard thresholds in the PGRE Percentiles.}
\label{fig:thresh}
\end{figure}

\begin{deluxetable}{cccccccccc}
\tabletypesize{\scriptsize}
\tablecolumns{10} 
\tablewidth{0pt}
 \tablecaption{Number of prize fellowship recipients who would have been eliminated by strict percentile thresholds in PGRE admission standards.}
 \tablehead{
 \colhead{} & \multicolumn{9}{c}{\textbf{Number (and percentage) of fellows below Physics GRE percentile threshold}} \\
 \colhead{$\mathbf{N_{total}}$} & \colhead{${<}10\%$} & \colhead{${<}20\%$} & \colhead{${<}30\%$} & \colhead{${<}40\%$} & \colhead{${<}50\%$} & \colhead{${<}60\%$} & \colhead{${<}70\%$} & \colhead{${<}80\%$} & \colhead{${<}90\%$}} 
 \startdata 
 \sidehead{\bf All}
149.0 &  4 (2.7\%) &  10 (6.7\%) &  16 (10.7\%) &  32 (21.5\%) &  44 (29.5\%) &  65 (43.6\%) &  85 (57.0\%) &  104 (69.8\%) &  121 (81.2\%) \\
 \sidehead{\bf Male}
75.0 &  1 (1.3\%) &  3 (4.0\%) &  6 (8.0\%) &  13 (17.3\%) &  15 (20.0\%) &  25 (33.3\%) &  34 (45.3\%) &  47 (62.7\%) &  59 (78.7\%)  \\
 \sidehead{\bf Female}
53.0 &  1 (1.9\%) &  4 (7.5\%) &  6 (11.3\%) &  14 (26.4\%) &  22 (41.5\%) &  32 (60.4\%) &  42 (79.2\%) &  47 (88.7\%) &  49 (92.5\%) \\
\sidehead{\bf Not Specified/Other} 
21.0 &  2 (9.5\%) &  3 (14.3\%) &  4 (19.0\%) &  5 (23.8\%) &  7 (33.3\%) &  8 (38.1\%) &  9 (42.9\%) &  10 (47.6\%) &  13 (61.9\%) \\  \enddata
\end{deluxetable}

\begin{figure}[h]
\centering
\includegraphics[width=\textwidth]{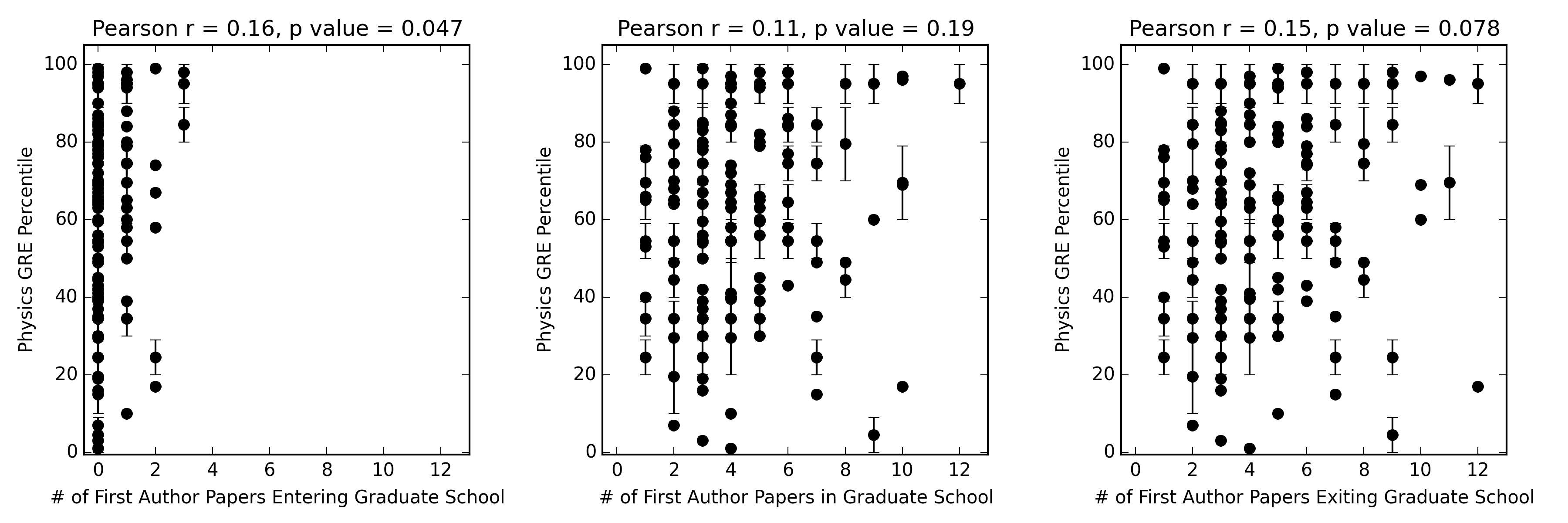}
\caption{Number of \textit{first author} publications written by fellows before graduate school (left panel), written in grad school (center panel), and total number of papers exiting graduate school (right panel) versus PGRE percentile score. Pearson $r$ statistics and corresponding $p$ values are cited at the top of each panel.}
\label{fig:gre_papers}
\end{figure}

\begin{figure}[h]
\centering
\includegraphics[width=0.45\textwidth]{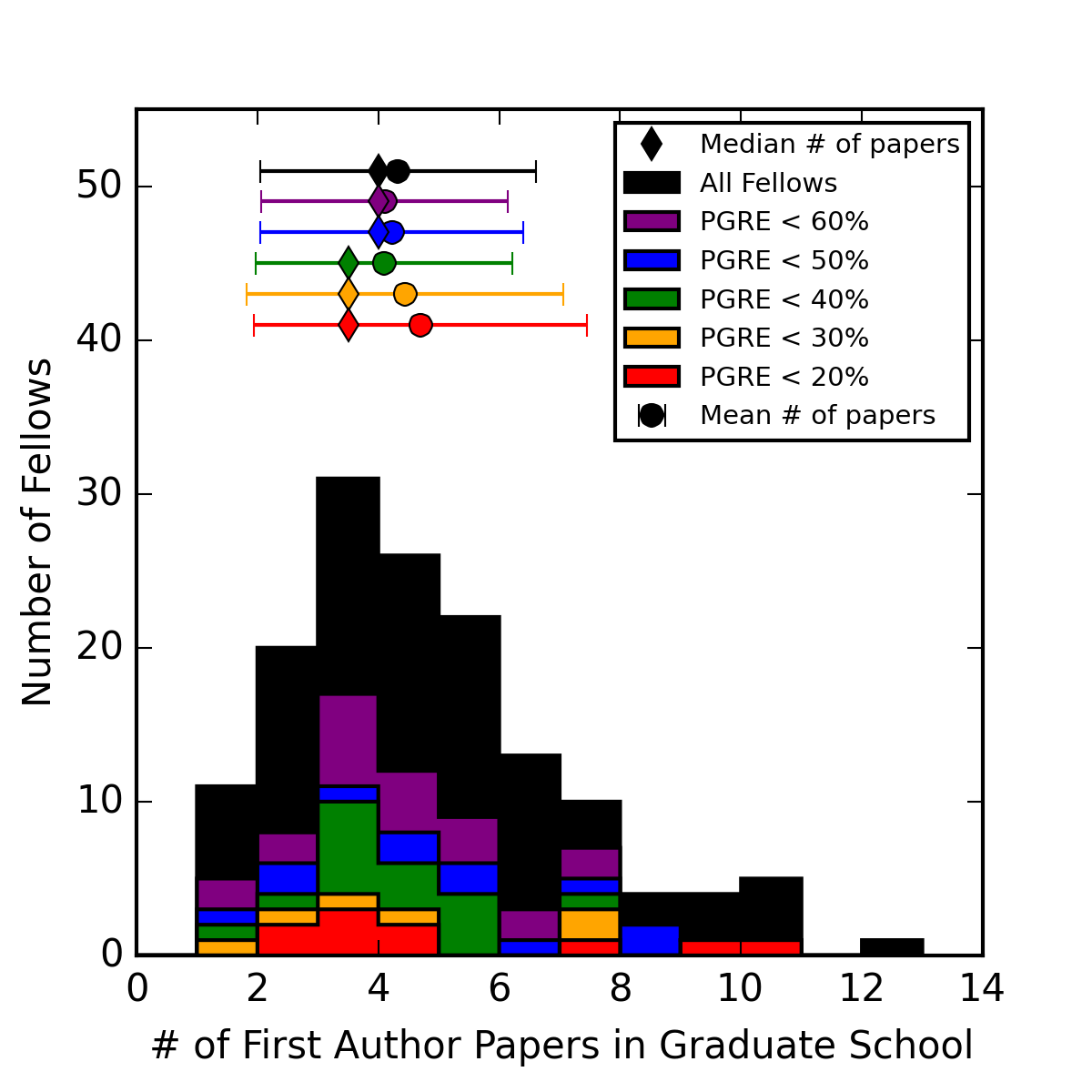}
\includegraphics[width=0.45\textwidth]{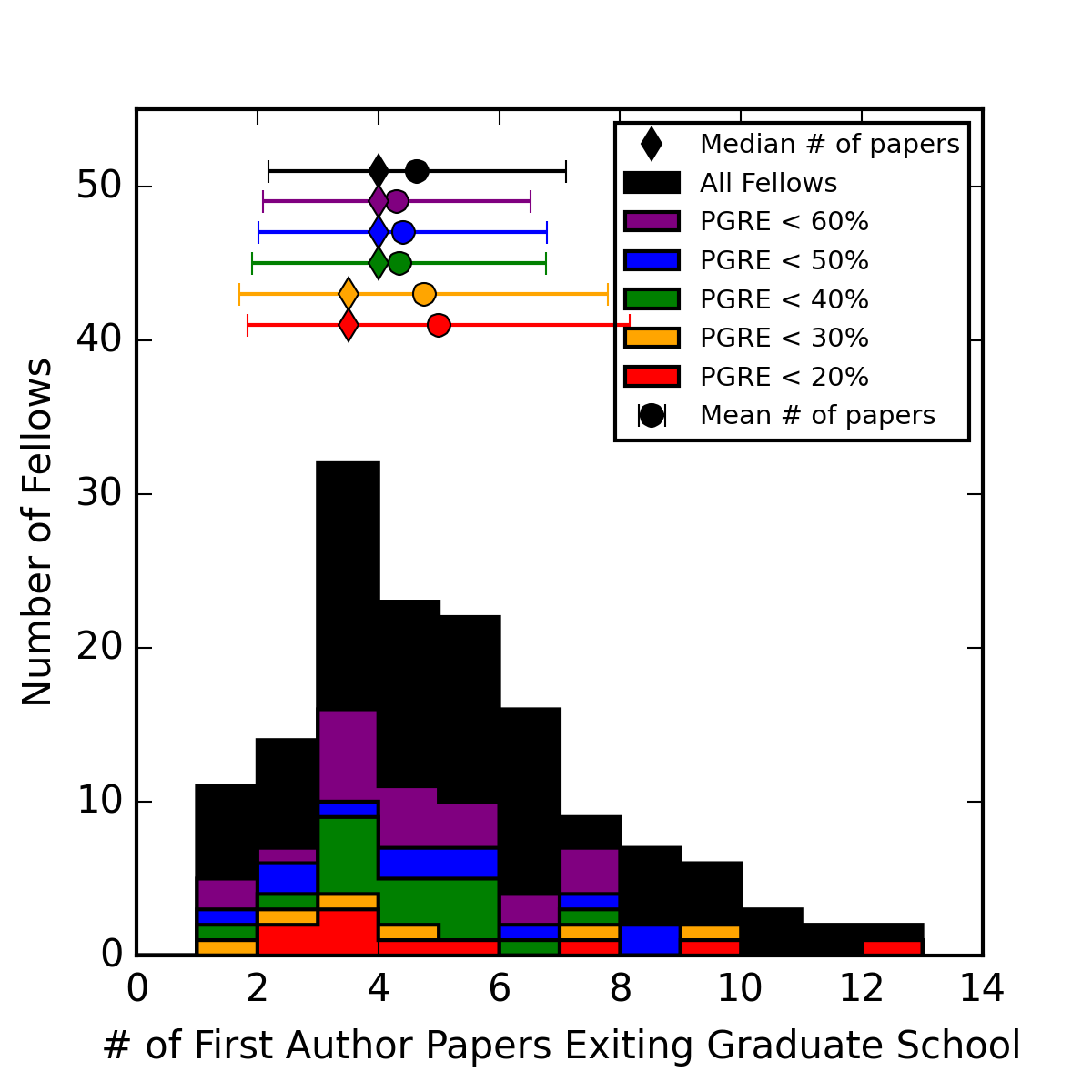}
\caption{Productivity of \textit{first-author} publications written by prize fellows during graduate school (left panel) and exiting graduate school (right panel). The mean and median number of first author publications for fellows below different PGRE thresholds are indicated by colored circles and diamonds, respectively - error bars indicate the standard deviation of paper counts in each subset. The general shape of the histogram and the average number of publications is insensitive to PGRE Score.}
\label{fig:paperhist}
\end{figure}

\begin{deluxetable}{cccccccccccc}
\tabletypesize{\scriptsize}
\tablecolumns{11} 
\tablewidth{0pt}
 \tablecaption{Total first author papers written by fellows below PGRE thresholds}
 \tablehead{
 \colhead{} & \colhead{} & \multicolumn{10}{c}{\textbf{Number of first author papers written by fellows below PGRE Threshold}} \\
 \colhead{} & \colhead{$\mathbf{N_{total}}$} & \colhead{${<}10\%$} & \colhead{${<}20\%$} & \colhead{${<}30\%$} & \colhead{${<}40\%$} & \colhead{${<}50\%$} & \colhead{${<}60\%$} & \colhead{${<}70\%$} & \colhead{${<}80\%$} & \colhead{${<}90\%$} & \colhead{All}} 
 \startdata 
All & 147 &  18 &  47 &  71 &  131 &  186 &  267 &  353 &  426 &  490 &  635  \\
Male & 74 &  9 &  16 &  31 &  61 &  70 &  109 &  153 &  201 &  244 &  341  \\
Female & 53 &  2 &  21 &  28 &  55 &  96 &  133 &  165 &  183 &  189 &  199  \\
Not Specified/Other & 20 &  7 &  10 &  12 &  15 &  20 &  25 &  35 &  42 &  57 &  95  \\
\enddata
\label{tab:Paperselim}
\end{deluxetable}

In addition to fellows' PGRE scores, we also considered whether PGRE score showed any correlation with first author publications; a statistically robust correlation would suggest that the PGRE may be indicative of research productivity in astronomy. Figure 4 compares the number of first author publications written by fellows at various stages of their career to PGRE scores. While there is marginal evidence ($p=0.047$) of a weak ($r=0.16$) correlation between PGRE score and number of papers published as an undergraduate, there is no statistically significant ($p<0.05$) correlation between PGRE score and total first author papers during or exiting graduate school among 147 prize fellows (2 of the 149 respondents who reported PGRE scores did not report first author paper statistics).

Similarly, Figure 5 compares histograms of first author papers written at various career stages for prize fellows whose PGRE scores fell below thresholds ranging from 20-60\%. The mean and median number of first author publications for fellows in each cutoff category are illustrated. The general shape of the histogram and the average number of publications are both insensitive to PGRE cutoff score.

Table 2 offers a numerical breakdown of these results, giving the total number of first author papers written by prize fellows with PGRE scores below thresholds ranging from 10-90\%.

\section{Discussion}
Based on the above results, we find no evidence that the PGRE can be used as an effective predictor of ``success" either in or beyond graduate school. The PGRE scores of individuals who have held prize fellowships do not adhere to any minimum cutoff, and show a significant deviation from the score distribution that would be expected if PGRE correlated with career performance. In addition, there is also no statistically significant correlation shown between PGRE score and the number of first author publications written by prize fellows either during or exiting graduate school. The average number of first author publications among prize fellows remains unchanged regardless of whether a PGRE cutoff score is applied.

Publication records also cannot simply be used to counteract low PGRE scores in admissions decisions; of the 15 fellows who scored $<$30\% on the PGRE, only 3 (20\%) had published one or more first author papers during their undergraduate careers. It is also worth noting that this is in agreement with the overall percentage of fellows who reported publishing one or more first author papers as undergraduates (34, or 23\%).

It is true, of course, that all the prize fellows included in this study were eventually admitted to a graduate program and successfully completed their PhDs. However, it is important to note that our sample includes prize fellows who received their PhDs overseas or from programs that do not require the PGRE.

While there are many factors at play in assessing a graduate school applicant pool (including hard-to-quantify elements such as CVs, reference letters, and statements of purpose) these data demonstrate that the use of PGRE scores to either eliminate or negatively impact graduate school applications is likely to remove a significant number of potentially successful astronomers from the applicant pool.

For example, we can consider a ``hard" cutoff of 60\% applied to a hypothetical applicant pool comprised of our poll respondents. Such a cutoff would have excluded 64 future prize fellowship
recipients. In addition, while this would exclude 43.6\% of the total applicant pool, the cutoff would disproportionately impact women, excluding 60.4\% of female applicants. Those prospective applicants would have gone on to write 267 (133) papers during their graduate careers.

Even a relatively low cutoff in PGRE scores would eliminate future prize fellows (4 at $<$10\%, 10 at $<$20\%). Overall, the data suggest that any application of a PGRE cutoff score in graduate admissions carries with it a substantial risk of eliminating potentially successful future astronomers from the applicant pool.

\acknowledgements{We acknowledge useful feedback and advice on this work from Caitlin Casey, Britt Lundgren, Elizabeth Mills, David Pitman, Andy Szymkowiak, Keivan Stassun, Meg Urry, and several anonymous respondents. We are grateful to all individuals who participated in our questionnaire.}

\end{document}